\documentclass[prd,amsmath,amssymb,twocolumn]{revtex4}
\usepackage{graphicx}
\usepackage{hyperref}
\usepackage{bm}
\usepackage{natbib}
\usepackage{tensor}

\def\be{\begin{equation}}
\def\ee{\end{equation}}
\def\ba{\begin{eqnarray}}
\def\ea{\end{eqnarray}}
\def\nn{\nonumber \\}

\def\GeV{{\rm \, GeV}}
\def\Mpc{{\rm \, Mpc}}
\def\TeV{{\rm \, TeV}}
\def\eV{{\rm \, eV}}
\def\meV{{\rm \, meV}}
\def\yr{{\rm \, yr}}

\def\cm{{\rm \, cm}}

\def\sec{{\rm \, sec}}
\def\sr{{\rm \, sr}}

\newcommand{\per}{\, .}
\newcommand{\com}{\, ,}
\newcommand{\rref}[1]{Ref.~\cite{#1}}
\newcommand{\fref}[1]{Fig.~\ref{#1}}
\newcommand{\eref}[1]{Eq.~(\ref{#1})}
\newcommand{\erefs}[2]{Eqs.~(\ref{#1})~and~(\ref{#2})}

\newcommand{\ax}{\varphi}
\newcommand{\gag}{g_{a\gamma}}

\newcommand{\Gauss}{\, {\rm G}}
\newcommand{\MeV}{\, {\rm MeV}}
\newcommand{\tx}{(t,{\bf x})}
\newcommand{\gm}{e_{m}}
\newcommand{\Lcal}{\mathcal{L}}
\newcommand{\Fcal}{\mathcal{F}}
\newcommand{\Hcal}{\mathcal{H}}
\newcommand{\QCD}{\text{\sc qcd}}

\newcommand{\kup}{k}

\newcommand{\hidden}[1]{}

\usepackage{epsfig}     
\usepackage{color}              
\usepackage{graphicx}   
\usepackage{hyperref}
\hypersetup{
    colorlinks=true,
    linkcolor=red,
    citecolor=blue,
}

\begin{document}
\title{
Implications of a Primordial Magnetic Field for Magnetic Monopoles, Axions, and Dirac Neutrinos
}

\date{\today}

\author{Andrew J. Long$^{a}$}
\email{andrewjlong@asu.edu}

\author{Tanmay Vachaspati$^{a,b}$}
\email{tvachasp@asu.edu}

\affiliation{ 
$^{a}$Physics Department, Arizona State University, Tempe, Arizona 85287, USA.\\
$^{b}$Physics Department and McDonnell Center for the Space Sciences,
Washington University, St. Louis, MO 63130, USA.
}


\begin{abstract}
We explore some particle physics implications of the growing evidence for a helical primordial magnetic
field (PMF). 
From the interactions of magnetic monopoles and the PMF, we derive an upper bound on the monopole number density, $n(t_0) < 1 \times 10^{-20} \cm^{-3}$, which is a ``primordial'' analog of the Parker 
bound for the survival of galactic magnetic fields.  
Our bound is weaker than existing constraints, but it is derived under 
independent assumptions. We also show how improved measurements of the PMF at different redshifts 
can lead to further constraints on magnetic monopoles. Axions interact with the PMF due to the 
$\gag \ax {\bf E}\cdot {\bf B}/4\pi$ interaction. Including the effects of the cosmological plasma, we find
that the helicity of the PMF is a source for the axion field. Although the magnitude of the source
is small for the PMF, it could potentially be of interest in astrophysical environments.
Earlier derived constraints from the resonant conversion of cosmic microwave background (CMB)
photons into axions lead to $\gag \lesssim10^{-9}~{\rm GeV}^{-1} $ for the suggested PMF strength
 $\sim 10^{-14} \Gauss$ and coherence length $\sim 10~{\rm Mpc}$. 
Finally we apply constraints on the neutrino magnetic dipole moment that arise from requiring successful
big bang nucleosynthesis in the presence of a PMF and we find $\mu_\nu \lesssim 10^{-16} \mu_B$.
%
\end{abstract}

\maketitle

\section{Introduction}
\label{sec:intro}

There is growing evidence for the existence of an intergalactic magnetic field from the observation of high energy gamma rays.  
It is likely that a magnetic field in intergalactic space would have been created in the early universe, since astrophysics alone is not expected to generate fields on such large length scales.  
(For a recent review on cosmic magnetic fields, see \rref{Durrer:2013pga}.) 
The discovery of a primordial magnetic field (PMF) has important ramifications for cosmology as it allows one to test models of magnetogenesis, which are often tied to the physics of inflation \cite{Turner:1987bw} cosmological phase transitions \cite{Vachaspati:1991nm, Ahonen:1997wh}, and baryogenesis \cite{Cornwall:1997ms, Vachaspati:2001nb, Long:2013tha}.  
The presence of a PMF after cosmological recombination can also aid in the formation of first stars \cite{1987QJRAS..28..197R} and provide the seed field for the galactic dynamo \cite{Ruzmaikin:book}.  
Additionally, the existence of a PMF in our universe opens the opportunity to place indirect constraints on exotic particle physics models where the new physics couples to electromagnetism.  
In this paper we will investigate the consequences of a PMF for models that contain magnetic monopoles, axions, and Dirac neutrinos with a magnetic moment.

Blazars that emit TeV gamma rays are expected to produce an electromagnetic cascade of lower energy gamma rays due to electron-positron pair production and the subsequent inverse Compton up-scattering of cosmic microwave background (CMB) photons \cite{1994ApJ...423L...5A, 1995Natur.374..430P, Neronov:2006hc, Elyiv:2009bx, Dolag:2009iv}.
In the presence of an intergalactic magnetic field, electrons and positrons directed toward the Earth can be deflected off of the line of sight, and those that are directed away from the Earth can be deflected back toward the line of sight.  
As a result the point source flux is depleted in the GeV band, and the blazar acquires a halo of GeV gamma rays.  
The non-observation of these GeV gamma rays was used to place a lower bound on the  magnetic field 
strength at the level of $B \gtrsim 10^{-16} \Gauss$ \cite{Neronov:1900zz, Tavecchio:2010mk, Dolag:2010ni}.  
This bound depends on modeling of the blazar flux stability and also the plasma instabilities during propagation,
and it may weaken substantially depending on these 
assumptions \cite{Dermer:2010mm, Broderick:2011av, Miniati:2012ge, Schlickeiser:2012xx}.  
The search for the GeV halo extended emission has been ongoing \cite{Ando:2010rb, Neronov:2010bi, Aleksic:2010gi, Alonso:2014ewa, Abramowski:2014uta}. 
Most recently, Chen et al. \cite{Chen:2014rsa} have found evidence for the halo in a stacked analysis of known blazars at $\sim 1~{\rm GeV}$ energies and interpret it to be due to a field with strength $B \sim 10^{-17}-10^{-15} \Gauss$.
For reference, measurements of the cosmic microwave background place an upper bound on the magnetic field strength at the level of $B \lesssim 10^{-9} \Gauss$ \cite{Ade:2015cva}.  

There are theoretical motivations for considering the possibility that the PMF is {\it helical}, {\it i.e.} there is an excess of power in either right- or left-circular polarization modes.  
Helical magnetic fields emerge in many models of magnetogenesis \cite{Caprini:2014mja, Cornwall:1997ms, Vachaspati:2001nb, Long:2013tha}, and helicity conservation dramatically impacts the evolution of the PMF, aiding in its survival and growth \cite{Kahniashvili:2012uj}.  
Recently, Tashiro et al. \cite{Tashiro:2013ita, Chen:2014qva} analyzed the diffuse gamma ray sky at 10-60~GeV energies to look for the parity-violating signature \cite{Tashiro:2013bxa, Tashiro:2014gfa} of a helical magnetic field.  
They find evidence for an intergalactic magnetic field with strength $B \sim 10^{-14} \Gauss$ on coherence scales $\lambda_B \sim 10 \Mpc$ and with left-handed helicity.
Although the results of Refs.~\cite{Tashiro:2013ita, Chen:2014qva} and \cite{Chen:2014rsa} appear inconsistent, it may be possible to reconcile them by noting that the weak bending approximation breaks down for $B \sim 10^{-14} \Gauss$ for gamma rays at $\sim 1~{\rm GeV}$ energies \cite{Chowdhury:InPrep}.  

Motivated by these recent results, we investigate if the existence of a helical PMF can be used to constrain other particle physics ideas in a cosmological setting.  
Our analysis is sufficiently general that our results will remain relevant even if the gamma ray observation results should change or go away, {\it e.g.} with more data.
However, for the purpose of numerical estimates we will use $B \sim 10^{-14} \Gauss$ and 
$\lambda_B \sim 10 \Mpc$ as the fiducial field strength and coherence length scale, and
we will take the magnetic field to have maximal (left-handed) helicity.

In Sec.~\ref{sec:monopoles} we consider the interaction of a hypothetical abundance of cosmic 
magnetic monopoles and a PMF.  
The magnetic field does work on the monopoles, and its field strength is thereby depleted.  
The constraints we obtain in this way are generally weaker than existing bounds but are obtained under a different set of assumptions.  
These results are summarized in Fig.~\ref{fig:constraints}.
We also discuss how heavy magnetic monopoles can lead to anomalous scaling of the energy density in the PMF.  
As observations of the PMF improve, they can be sensitive to the anomalous scaling and thus become a tool for further constraining magnetic monopoles.

In Sec.~\ref{sec:axions} we consider the interaction of an axion ($\ax$) with the PMF through the coupling $g_{a\gamma} \ax F {\tilde F}$.  
In this analysis we include the cosmological plasma, and thus we study the equations of magnetohydrodynamics coupled to an axion.  
Although we find that the axion has a negligible effect on the spectrum and evolution of the PMF, it is interesting to note that this conclusion is not sensitive to the assumed scale of Peccei-Quinn symmetry breaking, $f_{a}$, as long as $\gag \propto 1 / f_{a}$.  
In turn, the PMF leaves the evolution of the axion condensate largely unaffected.  
In principle the PMF damps the axion oscillations and the helicity of the PMF shifts the equilibrium point, but these effects are quadratic in the already-small magnetic field strength.  

In Sec.~\ref{sec:neutrinos} we consider the interaction of the neutrino magnetic dipole moment, $\mu_\nu$, with the PMF.  
Enqvist et al. \cite{Enqvist:1992di,Enqvist:1994mb} have shown that this interaction induces a spin-flip transition, which cannot be in equilibrium in the early universe without running afoul of constraints on the number of relativistic neutrino species.  
Using their result with our fiducial value of $B \sim 10^{-14} \Gauss$, we evaluate an upper bound on $\mu_{\nu}$, which is shown in \fref{fig:mag_mom}.  

We work in the {\rm CGS} system with $\hbar = c = 1$.  The unit of electric charge is 
$e = \sqrt{\alpha \hidden{\hbar c}} \simeq 0.085 \hidden{\sqrt{\hbar c}}$ with $\alpha \simeq 1 / 137$ the fine structure constant, 
and the unit of magnetic charge is 
$\gm = \hidden{\hbar c \times} 1 / 2e \simeq 5.9 \hidden{\sqrt{\hbar c}}$.  
The magnetic field is measured in Gauss, and $1 \Gauss \simeq 6.93 \times 10^{-20} \GeV^2 \hidden{(\hbar c)^{-3/2}}$.  
The reduced Planck mass is denoted by $M_P \simeq 2.4 \times 10^{18} \GeV \hidden{/c^2}$.  
The metric signature is $(+---)$, and the antisymmetric tensor normalization is $\epsilon^{0123} = +1$.  

\section{Magnetic monopoles}
\label{sec:monopoles}

A conservative cosmological bound on the energy density of magnetic monopoles is $\Omega_m \equiv \rho_m/\rho_{\rm cr} < 0.3$ where $\rho_m$ is the energy density in monopoles and $\rho_{\rm cr} \simeq 10^{-29} \, {\rm gm} \, c^2 / \cm^3$ is the critical cosmological energy density.  
The number density of nonrelativistic monopoles is $n_m = \rho_{m} / m \hidden{c^2}$ with $m$ the monopole mass, and the cosmological bound implies 
\begin{align}\label{eq:n0_cosmo}
	n_m < 0.3 \frac{\rho_{\rm cr}}{m \hidden{c^2} } \simeq (2 \times 10^{-23} \cm^{-3}) \left( \frac{m \hidden{c^2}}{10^{17} \GeV} \right)^{-1} \per
\end{align}
The bound grows weaker for lighter monopoles since they contribute less to the energy density for the same number density.  

The existence of the galactic magnetic field leads to another indirect bound.  
Magnetic monopoles tend to deplete a magnetic field in the same way that free electrons short out a conductor.  
The survival of the micro-Gauss galactic magnetic field implies an upper bound on the directed flux, $\Fcal$,
of magnetic charge onto the Milky Way.  
Requiring that the time scale for B-field depletion is longer than the dynamo time scale of B-field regeneration ($\tau_{\rm dyn} \simeq 10^{8} \yr$), leads to the so called Parker bound \cite{Parker:1970xv} 
\begin{align}
	\Fcal < 0.9 \times 10^{-16} \gm \cm^{-2} \sec^{-1} \sr^{-1} \per
\end{align}
Assuming that monopoles have unit charge and travel with velocity $v$, the Parker bound can be expressed as an upper bound on the monopole number density:
\begin{align}\label{eq:Parker_bound}
	n_m \approx \frac{(4 \pi \, \sr) \, \Fcal}{e_m \, v} < (4 \times 10^{-23} \cm^{-3}) \left( \frac{v}{10^{-3} c} \right)^{-1}
	\per
\end{align}
Just as the Parker bound is predicated on the existence of a magnetic field in the Milky Way, we expect that a similar bound can be inferred from the existence of a primordial magnetic field in the early universe.  

We study a gas of monopoles and antimonopoles immersed in a magnetic field that permeates the cosmological medium.  
The monopoles have mass $m$, magnetic charge $\gm$, and they are homogeneously distributed with number density $n_{m}(t)$.  
The magnetic field ${\bf B}\tx$ induces a Lorentz force of 
\be\label{eq:FB_def}
	{\bf F}_{\rm B} = \gm {\bf B} 
\ee
on a monopole at $\tx$, and it begins to drift along the field line with a velocity ${\bf v}$.  
The field does work by pushing the monopole, and in this way the monopole extracts energy from the magnetic field at a rate 
\begin{align}\label{eq:EBdot}
	\dot{E}_{\rm B} = {\bf F}_{\rm B} \cdot {\bf v} \per
\end{align}
To solve for the evolution of the magnetic field strength we must know the monopole velocity.  
Prior to electron-positron annihilation, the monopole's velocity is restricted by elastic scattering with the cosmological medium, but afterward it can free stream.  
We will consider each of these cases in turn.

\subsection{Friction Dominated Regime}

In the epoch prior to $e^+ e^-$ annihilation the cosmological medium was dense with electromagnetically charged particles.  
In this regime, monopoles interact with the medium through elastic scattering such as  $M + e^{\pm} \to M + e^{\pm}$ with $M$ the monopole.  
It is safe to assume that the monopole's rest mass is much larger than the kinetic energy of particles in the plasma.  
This allows us to characterize the effective interaction with a drag force that takes the form \footnote{We obtain identical results when characterizing the interaction with a magnetic analog of Ohm's law, and the conductivity is taken to be $\sigma_{M} = \gm^2 n_{m} / f_{\rm drag}$.  }
\be
	{\bf F}_{\rm drag} = - f_{\rm drag} {\bf v} \per
\ee
At the time of interest, the scatterers are relativistic with energy comparable to the temperature $T$ of the plasma, and they are in thermal equilibrium with number density $n \sim T^3 \hidden{(\hbar c)^{-3}}$.  
For such a system the drag coefficient takes the form \cite{Goldman:1980sn, VilenkinShellard:1994}
\begin{align}
	f_{\rm drag} \approx \beta \, e^2 \gm^2 \, g_{em} T^2 \hidden{(\hbar^{-3} c^{-4})}
\end{align}
where $g_{em}(t)$ is the number of relativistic, charged degrees of freedom in thermal equilibrium at time $t$ and $\beta$ is an $O(1)$ number related to the spin character and charge of the scatterers.  
We will drop $\beta$ from this point onward since it is parametrically redundant with $g_{em}$.  
Also note that $g_{em}$ and $T$ depend on time, but this can be ignored on short time scales as compared to the Hubble time scale.

The monopole's equation of motion is  
\be
	m \, \dot{\bf v} = \gm {\bf B}  - f_{\rm drag} {\bf v} \per
	\label{eq:dotv}
\ee
We assume that the distance traveled by the monopole is small compared to the correlation length 
$\lambda_B$ of the magnetic field, and we treat ${\bf B}$ as uniform. \eref{eq:dotv} immediately
gives the terminal velocity of the monopoles,
\be\label{eq:vterm_def}
	{\bf v}_{\rm term} = \frac{\gm {\bf B}}{e^2 \gm^2 g_{em} T^2} \hidden{\hbar^{3} c^{4}} \com
\ee
which is achieved on a time scale
\be\label{eq:tauterm_def}
	\tau_{\rm term} = \frac{m}{e^2 \gm^2 g_{em} T^2} \hidden{\hbar^{3} c^{4}} \per
\ee
Comparing with the Hubble time $t_H \sim M_P / T^2$ (radiation era) we have $\tau_{\rm term} \ll t_H$ provided that $m < g_{em} M_P$, and thus the cosmological expansion is negligible.  

At the present cosmic epoch, the photon temperature is $\sim 10^{-4}~{\rm eV}$ and $B \sim 10^{-14} \Gauss$.
Assuming $B \propto T^2$, we get $B \simeq 10^{6} \Gauss$ at $T \simeq \MeV$ when $g_{em} \simeq 10$. 
These estimates give $v_{\rm term} \simeq 10^{-8} \hidden{c}$, which validates our uses of the 
non-relativistic equation of motion. As a consequence, the distance traveled by a monopole during
$\tau_{\rm term}$ is quite small, $d_{\rm term} <10^{-8} \tau_{\rm term}$. We shall assume that the 
correlation length of the magnetic field is larger, $\lambda_B > d_{\rm term}$, thus justifying our
treatment of the magnetic field as being uniform.

The magnetic field's response to this current is given by the magnetic analog of Ampere's law, 
\begin{align}
\label{eq:dotB}
	\dot{\bf B} = - 4 \pi \, {\bf j}_{M} = - 4\pi \gm n_m {\bf v} \per
\end{align}
where $n_m$ is the number density of monopoles (assumed equal to the number density of antimonopoles).
Here we have used ${\bf E}=0$ since electric fields are screened due to the high electrical conductivity of the cosmological medium.  
For typical parameters, the inter-monopole spacing is small compared to the correlation length of the magnetic field, $n_m^{-1/3} \ll \lambda_B$, and we can interpret $n_m$ and ${\bf B}$ as coarse grained quantities on this length scale. 
Then we insert ${\bm v}={\bm v}_{\rm term}$ from 
\eref{eq:vterm_def} into \eref{eq:dotB} to get the solution,
\begin{align}
	{\bf B}(t) = {\bf B}(t_i) \, e^{- (t-t_i) / \tau_{\rm decay}}
\end{align}
where the decay time scale of the magnetic field is given by
\be\label{eq:taudecay_def}
	\tau_{\rm decay} = \frac{e^2 \gm^2 g_{em} T^2}{4 \pi \gm^2 n_m} \hidden{\hbar^{-3} c^{-4}} .
\ee
In obtaining this solution, we have assumed ${\bm v}={\bm v}_{\rm term}$ which is justified if
the monopoles reach terminal velocity much more quickly than the decay time scale, 
{\it i.e.} $\tau_{\rm decay} \gg \tau_{\rm term}$. As we will see below, this condition is satisfied for the
range of parameters of interest to us.

To ensure survival of the magnetic field, we require that $\tau_{\rm decay}$ is much larger than the Hubble time
at temperature $T$, 
\be
	t_{H} = \frac{1}{2H} \simeq 1.5 \frac{M_P}{g_{\ast} T^2} \hidden{\hbar c^2}
	\label{eq:tautH}
\ee
with $H$ the Hubble parameter and $g_{\ast}$ the effective number of relativistic degrees of freedom.  
Substitution of \eref{eq:taudecay_def} into $\tau_{\rm decay} > t_{H}$ now leads to a constraint on the number density of monopoles, 
\be 
n_m < \frac{e^2 \gm^2}{6 \pi \gm^2} \frac{g_{em} g_{\ast} T^4}{M_P} \hidden{(\hbar^{-4} c^{-6})} \com
\ee
when the universe had temperature $T$.  

 \begin{figure}[t]
\hspace{0pt}
\vspace{-0in}
\begin{center}
\includegraphics[width=0.47\textwidth]{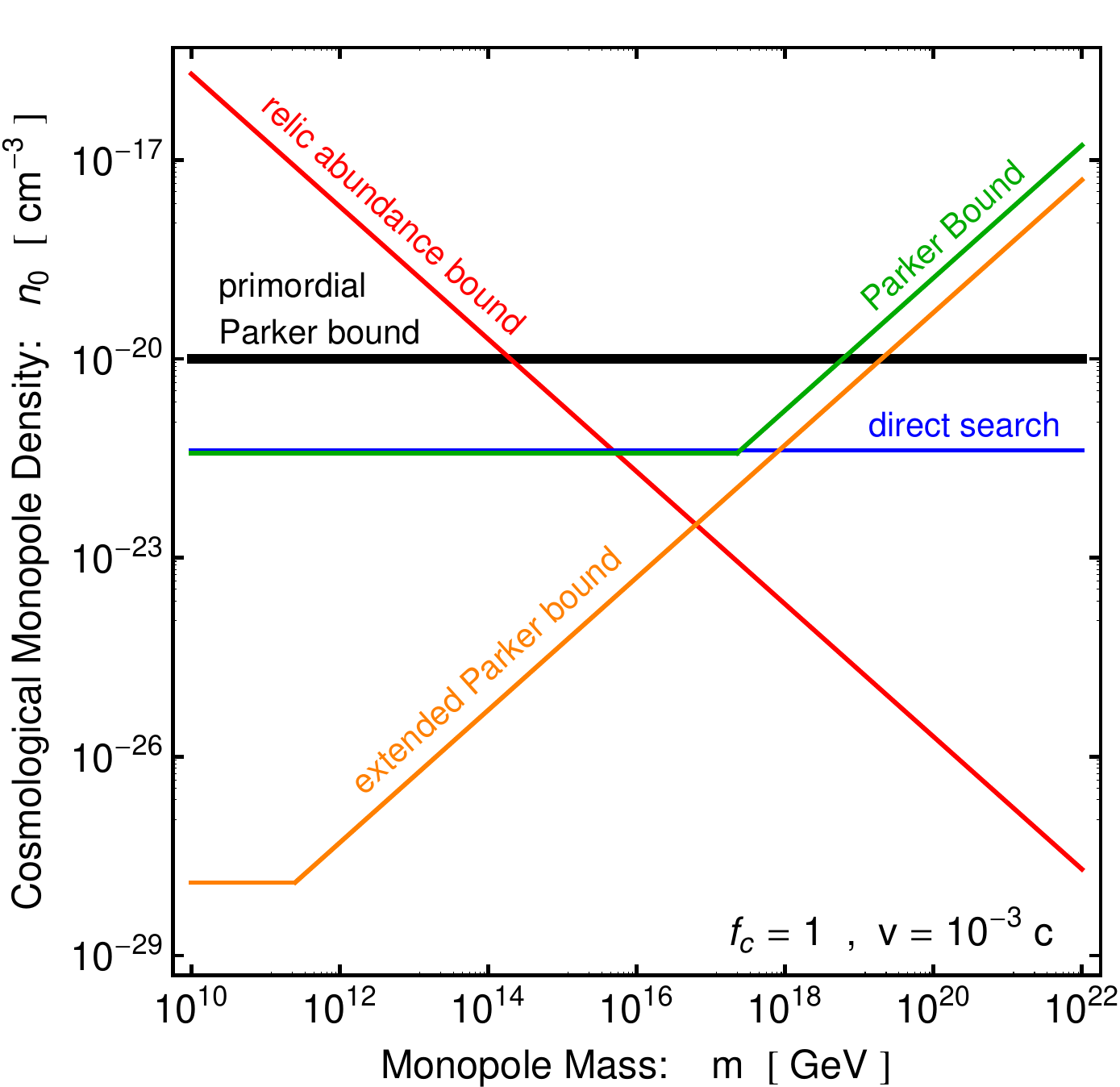} 
\caption{
\label{fig:constraints}
A summary of upper bounds on the magnetic monopole abundance from this work and the literature.  
{\it Black:} the requirement of survival of the primordial magnetic field (``primordial Parker bound''), derived here in \eref{eq:n0_bound}.  
{\it Red:}  the cosmological abundance bound in \eref{eq:n0_cosmo}.  
{\it Blue:}  direct search constraints \cite{Agashe:2014kda} (see the magnetic monopole review). 
{\it Green:}  the requirement of survival of the Galactic magnetic field (Parker bound), given by \rref{Turner:1982ag}.  
{\it Orange:}  the requirement of survival of the Galactic seed field (``extended Parker bound''), given by \rref{Adams:1993fj}.  
We take $v \simeq 10^{-3} \hidden{c}$ and assume that monopoles are unclustered, $f_{c} \simeq 1$.  
If the monopoles are clustered then the Parker bound, extended Parker bound, and direct search limits move down by a factor of $f_{c} \sim 10^{5}$.
}
\end{center}
\end{figure}

The strongest bound is obtained when $T$ is smallest.  
Since our calculation assumes that monopoles scatter on relativistic, charged particles with a thermal abundance, the last time at which this is possible is the epoch of $e^+ e^-$ annihilation.  
At this time $T_{\rm ann} \simeq 1 \MeV$, $g_{em} \approx g_{\ast} \approx g_{\ast S} \simeq 10.75$.  
To translate this into a bound on the monopole number density today, denoted by $n_0$, we multiply by 
$(a_{\rm ann}/a_0)^3 = (g_{\ast S, 0} T_0^3 / g_{\ast S} T_{\rm ann}^3)$ with $g_{\ast S,0} \simeq 3.91$ and 
$T_0 \simeq 2.3 \times 10^{-4} \eV$ the temperature of the microwave background photons today.  
We find an upper bound on the monopole number density today, 
\begin{align}\label{eq:n0_bound}
	n_0 & < \frac{e^2 \gm^2}{6\pi \gm^2} \frac{g_{\ast S, 0} g_{\ast} g_{em}}{g_{\ast S}} \frac{T_{\rm ann} T_0^3}{M_p} \hidden{(\hbar^{-4} c^{-6})} 
	\simeq 1 \times 10^{-20} \cm^{-3} \per
\end{align}
If this bound is not satisfied then any primordial magnetic field would have been exponentially depleted by the time of electron-positron annihilation.  
Due to the close connection with the Parker bound for survival of galactic magnetic fields, we will refer to \eref{eq:n0_bound} as the ``primordial Parker bound.''  

In \fref{fig:constraints} we compare the primordial Parker bound in \eref{eq:n0_bound} with other constraints derived previously in the literature.  
Since these constraints are typically expressed as a bound on the monopole flux, we translate into a bound on the number density using $(4 \pi \, {\rm sr}) \mathcal{F} \approx f_{c} n v$.  
Here $v$ is the average monopole velocity and $f_{c} = n_{\rm galaxy} / n_{\rm cosmo.}$ is the enhancement factor that accounts for clustering of monopoles in the galaxy.  
For clustered monopoles $f_{c} \sim 10^{5}$, but otherwise $f_{c} \sim 1$.  
For the extended Parker bound calculation of \rref{Adams:1993fj}, we take $B_{\rm seed} = 10^{-11} \Gauss$.
For the direct search constraint we show a relatively conservative and robust limit of $\Fcal < 10^{-15} \cm^{-2} \sec^{-1} \sr^{-1}$, but stronger constraints are available for specific monopole parameters \cite{Agashe:2014kda}.  
From the figure, one can see that the primordial Parker bound becomes stronger than the cosmological bound for light monopoles, $m \lesssim 5 \times 10^{13} \GeV \hidden{/c^2}$, but it always remains weaker than the direct search constraints.

\subsection{Free Streaming Regime}

After cosmological electron-positron annihilation the number density of these scatterers decreases by a factor of $\sim 10^{-10}$.  
The monopoles experience very little drag force, and they can be accelerated freely by the magnetic field.  
For a uniform and static magnetic field, the solution of \eref{eq:dotv} with $f_{\rm drag} = 0$ is simply $v = e_{m} B t / m$, or for an inhomogenous field with domains of size $\lambda_{B}$ we find 
\be
v(t) \sim \frac{e_m B \lambda_B}{m} \sqrt{\frac{t}{\lambda_B}}
\label{eq:vmon}
\ee
if the motion is diffusive.  
The monopole becomes relativistic when $v(t_{\rm rel}) \sim 1 \hidden{\times c}$ and comparing this time with the present age of the universe gives 
\be
\frac{t_{\rm rel}}{t_0} \sim \left ( \frac{m}{e_m Bt_0} \right )^2 \frac{t_0}{\lambda_B}
\sim 10^{18} \left ( \frac{m}{M_P} \right )^2 \frac{t_0}{\lambda_B}.
\ee
With $\lambda_B \sim {\rm Mpc}$ and $t_0 \sim 10 \, {\rm Gpc}$, we find that monopoles are relativistic today if $m \lesssim 10^8 \GeV$, and they are non-relativistic otherwise.

The above estimate ignores backreaction of the monopoles on the PMF. 
To check for consistency, we compare the kinetic energy in monopoles $\rho_{\rm kin}$ to the
energy density available in the PMF $\rho_{B} = B^2 / 8 \pi$.  
For relativistic monopoles we should have $\rho_B > \rho_{\rm kin} \gg m \, n_m$, and this provides an upper bound on the number density of monopoles for which the velocity estimate in \eref{eq:vmon} can be expected to hold.  
Taking $B \sim 10^{-14} \Gauss$ we find 
\be
	n_m \ll 10^{-35} \left ( \frac{10^8 ~{\rm GeV}}{m}\right ) ~ {\rm cm^{-3}} \per
\ee
Since are interested in much larger number densities, as indicated by the bound in \eref{eq:n0_bound}, we cannot use the velocity relation in \eref{eq:vmon}, but instead the monopole and magnetic field equations will need to be evolved simultaneously.

Without friction to provide a means of energy dissipation, the monopoles cannot deplete the magnetic field strength.  
Instead there is a conservative exchange of energy between the magnetic field and the kinetic energy of the monopoles.  
This co-evolution can lead to an anomalous departure from the usual power law scaling behavior of the magnetic field energy density if the monopoles are non-relativistic.  
This can be seen from the following argument.  
In the absence of the monopole gas, the energy density in the magnetic field redshifts like radiation $\rho_{B} \sim ( 1 + z)^{4}$ where $z$ is the cosmological redshift.  
Meanwhile the kinetic energy stored in a gas of non-relativistic particles redshifts more quickly.  
We can write the kinetic energy density as $\rho_{\rm kin} = n_m p^2/ (2m)$ where $p$ is the typical momentum and $n_m$ is the monopole number density.  
Since $p \sim (1+z)$ and $n_m \sim (1+z)^3$, the kinetic energy density redshifts like $\rho_{\rm kin} \sim (1+z)^{5}$.  
If energy is transferred quickly between the monopoles and magnetic field then we might expect $\rho_{B} \sim \rho_{\rm kin.} \sim (1+z)^{9/2}$ where $9/2$ is the average of $4$ and $5$. 
This result is confirmed by the full calculation, which we will now present.  

Once again we consider a gas of monopoles coupled to a magnetic field, but we now include the effects of cosmological expansion.  
Let $X^{\lambda}(\tau)$ be the world line of a monopole, and let $U^{\lambda}(\tau) = dX^{\lambda}/d\tau$ be its 4-velocity.  
The monopole equation of motion, given previously by \eref{eq:dotv}, is now replaced by
\begin{align}
	& m \left[ \frac{d U_{m}^{\lambda}}{d\tau} + \Gamma^{\lambda}_{\mu \nu} U_{m}^{\mu} U_{m}^{\nu}  \right] 
	= \gm \tensor{\widetilde{F}}{^{\lambda}_{\mu}} U_{m}^{\mu} \hidden{c^{-1}}
\end{align}
where $\Gamma^{\lambda}_{\mu \nu}$ is the Christoffel symbol.  
The magnetic analog of Ampere's law in \eref{eq:dotB} is now replaced by 
\begin{align}
	& \nabla_{\alpha} \widetilde{F}^{\alpha \beta} = 4\pi \hidden{c^{-1}} j_{M}^{\beta} 
\end{align}
where $\nabla_{\alpha}$ is the covariant derivative.  
The magnetic current density $j_{M}^{\mu}$ arises from monopoles with velocity $U_{m}^{\mu}$ and antimonopoles with velocity $U_{\bar{m}}^{\mu}$.  
It can be written as 
\begin{align}
	j_{M}^{\mu} & = \gm \left( n_{m} \, U_{m}^{\mu} - n_{\bar{m}} \, U_{\bar{m}}^{\mu} \right) \com
\end{align}
and it satisfies the conservation law 
\begin{align}
	& \nabla_{\mu} j_{M}^{\mu} = 0 \per
\end{align}
The spatial component of the 4-velocity is the {\it comoving} peculiar velocity $({\bf U}_{m})^{i} = U_{m}^{i}$, and with an additional factor of $a$ we form the ${\it physical}$ peculiar velocity $({\bf v}_{m})^{i} = a U_{m}^{i}$.  
Neglecting the electric field and spatial gradients, the system of equations can be put into the form
\begin{subequations}\label{eq:coevolution}
\begin{align}
	& \partial_{\eta} (a {\bf v}_{m}) = \frac{\gm}{m} (a^2 {\bf B}) \\
	& a \partial_{\eta} (a^2 {\bf B}) = - 4\pi \gm (a^3 n_{m}) \, (a {\bf v}_{m}) \\
	& \partial_{\eta} \left( a^3 n_{m} \right) = 0 
\end{align}
\end{subequations}
where $d\eta = dt / a = da / Ha^2$ is the conformal time coordinate.  
The third equation implies that the number of monopoles per comoving volume is conserved; $n_0 = a^3 n_{m}$ is the number density of monopoles today.  
Then if not for the additional factor of $a$ in the second equation, the solutions would simply be oscillatory with angular frequency 
\begin{align}
	\omega_{\rm pl} = \sqrt{ \frac{4\pi \gm^2 n_0}{m}} \per
\end{align}
This is just the usual formula for plasma frequency, but instead of electron charge, mass, and density, here we find the corresponding parameters for the monopole gas.  

To solve these equations we must relate $a$ to $\eta$.  
During the radiation era we have $\eta = \eta_{i} + (a-a_{i}) / H_{i} a_{i}^2 \approx a / H_{i} a_{i}^2$, and the solution is 
\begin{subequations}\label{eq:Rera_B_and_v}
\begin{align}
	{\bf B} & = 
	\left( \frac{a}{a_{i}} \right)^{-9/4} \left[ \frac{ J_{0}(\phi) \ {\bf B}_{1}
	+ Y_{0}(\phi) \ {\bf B}_{2} }{ (a/a_{i})^{-1/4} } \right]
	\\
	{\bf v}_{m} & = 
	\left( \frac{a}{a_{i}} \right)^{-3/4} \left[ \frac{ \phi \ J_{1}(\phi) \ {\bf v}_{1}
	+ \phi \ Y_{1}(\phi) \ {\bf v}_{2} }{(a/a_{i})^{1/4}} \right] 
\end{align}
\end{subequations}
where
\begin{align}
	\phi \equiv \frac{ 2 \tilde{\omega} \, \eta_{i}}{a_{i}} \sqrt{a} \per
\end{align}
At late times $\phi \gg 1$, and all of the Bessel functions go to zero with an envelop $\sim \phi^{-1/2} \sim a^{-1/4}$.  
Then the terms in square brackets do not scale with $a$ and we find $B \sim a^{-9/4}$ and $v_{m} \sim a^{-3/4}$.  
One can check that the energy densities scale in the same way 
\begin{subequations}\label{eq:anom_scaling}
\begin{align}
	& \rho_{B} = \frac{|{\bf B}|^2}{8\pi} \propto a^{-9/2} \\
	& \rho_{\rm kin.} = \frac{m}{2} |{\bf v}_{m}|^2 \, n_{m}  \propto a^{-9/2} \com
\end{align}
\end{subequations}
which confirms our earlier argument.  

During the matter era we have $\eta = \eta_{i} + 2 (\sqrt{a} - \sqrt{a_i}) / H_i a_i^{3/2} \approx 2 \sqrt{a} / H_i a_{i}^{3/2}$, and the solution is 
\begin{subequations}\label{eq:Mera_B_and_v}
\begin{align}
	{\bf B} & = 
	\left( \frac{a}{a_i} \right)^{-9/4} 
	\sum_{s=\pm1} \left( \frac{a}{a_i} \right)^{\frac{i}{4} s \sqrt{ 4 \tilde{\omega}^2 \eta_i^2 / a_i - 1} } \! {\bf B}_{s} 	\\
	{\bf v}_{m} & = 
	\left( \frac{a}{a_i} \right)^{-3/4} 
	\sum_{s=\pm1} \left( \frac{a}{a_i} \right)^{\frac{i}{4} s \sqrt{ 4 \tilde{\omega}^2 \eta_i^2 / a_i - 1} } \! {\bf v}_{s} 	\per
\end{align}
\end{subequations}
For typical parameters we have $4 \tilde{\omega}^2 \eta_{i}^2 / a_{i} \gg 1$, and the solution is oscillatory with a power law envelope.  
We find the same anomalous scaling as in the radiation era, {\it cf.} \eref{eq:anom_scaling}.  
If we remove the monopoles from the problem by sending $n_{m} , \tilde{\omega} \to 0$ then the would-be oscillatory factors become a power law decay, and we regain the usual scaling $B \sim a^{-2}$ and $v_{m} \sim a^{-1}$.  

This anomalous scaling does not provide constraints on the monopole number density.  
However, it does affect the way that we translate constraints on the magnetic field in the early universe into the value of the magnetic field today.  
Measurements of the cosmic microwave background restrict the magnetic field energy density to be less than $\sim 10^{-5}$ of the photon energy density at the time of recombination \cite{Ade:2015cva}
\begin{align}
	\rho_{B}(z_{\rm rec}) \lesssim 10^{-5} \rho_{\gamma}(z_{\rm rec}) \per
\end{align}
Using the scaling relations, $\rho_{B} \sim (1+z)^{9/2}$ and $\rho_{\gamma} \sim (1+z)^{4}$ this inequality implies that the magnetic field energy density today is bounded by 
\begin{align}\label{eq:tighten_CMB}
	\rho_{B,0} & \lesssim 10^{-5} \rho_{\gamma,0} (1+z_{\rm rec})^{-1/2} \nn
	& \simeq (3 \times 10^{-10} \Gauss)^2 \left( \frac{1+z_{\rm rec}}{1300} \right)^{-1/2} 
\end{align}
where $\rho_{\gamma,0} \approx 2\pi^2 T_0^4/30$ is the CMB energy density today. 
If it were not for the anomalous redshifting, the constraint on the B-field strength would be weaker by a factor of $(1+z_{\rm rec})^{1/4} \simeq 6$. 

The anomalous scaling of the field strength in \eref{eq:anom_scaling} can become a tool in the future as measurements of the PMF improve.
By measuring the magnetic field strength at different redshifts, say by using TeV blazars at different distances, we can directly probe the anomalous scaling, and hence obtain a new handle on the relic density of non-relativistic magnetic monopoles.

\section{Axions}
\label{sec:axions}

Consider an axion $\ax(x)$ coupled to the electromagnetic field $A_{\mu}(x)$.  
The Lagrangian takes the form \cite{Kim:2008hd}
\begin{align}\label{axLag}
	\Lcal & = \frac{1}{2} (\partial_{\mu} \ax)^2 
	- \frac{m_a^2}{2} \ax^2 \hidden{\frac{c^2}{\hbar^2}} 
	- \frac{1}{16 \pi}F_{\mu\nu} F^{\mu\nu} 
	\nn & \quad  
	- \frac{\gag}{16\pi} \ax F_{\mu\nu} {\widetilde F}^{\mu\nu} 
	- \hidden{\frac{1}{c}} A_{\mu} j^{\mu}
\end{align}
where $\widetilde{F}^{\mu \nu} = \frac{1}{2} \epsilon^{\mu \nu \alpha \beta} F_{\alpha \beta}$ is the dual field strength tensor, and $j^{\mu} = (\hidden{c}\rho,{\bf j})$ is the electromagnetic current arising from the charged Standard Model fields.  
Our analysis is sufficiently general to apply to any axion or axion-like particle described by \eref{axLag}, but as a fiducial reference point we will consider a QCD axion with Peccei-Quinn scale of $f_a \simeq 10^{10} \GeV \hidden{/\sqrt{\hbar c}}$, an axion mass of $m_a \hidden{c^2} \approx \Lambda_{\QCD}^2 / f_a \simeq 1 \meV$, and a photon-axion coupling constant $\gag \approx \alpha / (2 \pi f_a) \simeq 10^{-13} \GeV^{-1} \hidden{\sqrt{\hbar c}}$.  

The classical axion condensate obeys the field equation 
\be
\Box \ax + m_a^2 \ax \hidden{\frac{c^2}{\hbar^2}}  = \frac{\gag}{4\pi} {\bf E} \cdot {\bf B}
\label{axeq}
\ee
where we have used $F_{\mu \nu} \widetilde{F}^{\mu \nu} = - 4 {\bf E} \cdot {\bf B}$.  
The electromagnetic field evolves according to the constraint equation
\begin{align}
	\partial_{\mu} \widetilde{F}^{\mu \nu} = 0
\end{align}
and the modified field equation 
\begin{align}
	\partial_{\mu} F^{\mu \nu} + \gag \, \partial_{\mu} \ax \widetilde{F}^{\mu \nu} = \frac{4\pi}{c} j^{\nu} \per
\end{align}
In terms of the electric and magnetic vector fields we have 
\begin{subequations}\label{eq:EM_eqns}
\begin{align}
	& {\bm \nabla} \cdot {\bf B} = 0 \label{divBeq} \\
	& {\bm \nabla}\times {\bf E} + \frac{1}{c} \frac{\partial {\bf B}}{\partial t} = 0 \label{dotBeq} \\
	& {\bm \nabla} \cdot {\bf E} + \gag {\bm \nabla} \ax \cdot {\bf B} = 4\pi \rho \label{divEeq} \\
	& {\bm \nabla} \times {\bf B} - \frac{1}{c} \frac{\partial {\bf E}}{\partial t} - \gag \dot{\ax} \hidden{\frac{1}{c}} {\bf B} - \gag {\bm \nabla} \ax \times {\bf E} = \frac{4\pi}{c} {\bf j} \label{dotEeq} \per
\end{align}
\end{subequations}
Note the presence of the additional terms arising from the spatio-temporal variation of the axion field.  

We seek to study the coevolution of the coupled axion and electromagnetic fields.  
\erefs{axeq}{eq:EM_eqns} describe a non-dissipative system.  
Dissipation is introduced as the electromagnetic field couples to charged particles in the cosmological medium, which opens an avenue for energy to be lost in the form of heat.  
This coupling is parametrized by the conductivity $\sigma$ which appears in Ohm's law
\be
	{\bm j} = \sigma \left ( {\bf E} + \hidden{\frac{1}{c}} {\bf v} \times {\bf B} \right ) 
\label{jeq}
\ee
where ${\bf v}\tx$ is the local velocity of the plasma.  
Prior to the epoch of $e^+ e^-$ annihilation, free charge carriers were abundant and the cosmological medium had a high conductivity \cite{Baym:1997gq}
\begin{align}\label{eq:sigma_def}
	\sigma \approx \hidden{\hbar^{-1}} T / \alpha 
\end{align}
where $\alpha \simeq 1/137$ is the fine structure constant.  
Ohm's law allows us to eliminate ${\bf j}$ and thereby reduce the system of equations in four unknowns $\{ {\bf E}, {\bf B}, {\bf j}, \ax \}$ to a set of equations describing only three unknowns:  
\begin{subequations}
\begin{align}
	& \ddot{\ax} - \hidden{c^2} \nabla^2 \ax + m_a^2 \ax \hidden{\frac{c^4}{\hbar^2}}  = \hidden{c^2} \frac{\gag}{4 \pi} {\bf E} \cdot {\bf B} \label{eq:eom1a} \\
	& \dot{{\bf B}} = - \hidden{c} \, {\bm \nabla} \times {\bf E} \label{eq:eom2a} \\
	& \dot{{\bf E}} = \hidden{c} \, {\bm \nabla} \times {\bf B} - \gag \dot{\ax} {\bf B} - \hidden{c} \gag {\bm \nabla} \ax \times {\bf E} \nn
	& \qquad - 4 \pi \sigma {\bf E} - 4 \pi \sigma {\bf v} \times {\bf B} \hidden{c^{-1}} \label{eq:eom3a} \per
\end{align}
\end{subequations}
In the MHD approximation (nonrelativistic flow) we can neglect the displacement current since it is negligible compared to the curl of the magnetic field, $|\dot{{\bf E}}| / \hidden{c} |{\bm \nabla} \times {\bf B}| \sim (v/c)^2 \ll 1$ \cite{Choudhuri:1998}.  
Then \eref{eq:eom3a} becomes algebraic in ${\bf E}$, and we can solve it to eliminate ${\bf E}$ from
the remaining equations.  
Focusing now on a homogenous axion field, the system of equations reduces to 
\begin{subequations}
\begin{align}
	& \ddot{\ax} + \gag^2 \frac{\eta_{\rm d} |{\bf B}|^2}{4 \pi} \dot{\ax} + m_a^2 \ax \hidden{\frac{c^4}{\hbar^2}}  = 
	\hidden{c} \frac{\gag \eta_{\rm d}}{4 \pi} {\bf B} \cdot {\bm \nabla} \times {\bf B}	
	 \label{axmhdeq} \\
	& \dot{{\bf B}} = 
	{\bm \nabla} \times \left( {\bf v} \times {\bf B} \right)
	+ \eta_{\rm d} \, \nabla^2 {\bf B} 
	+ \hidden{\frac{1}{c}} \gag \eta_{\rm d} \dot{\ax} \, {\bm \nabla} \times {\bf B}  
	\label{mhdeq}
\end{align}
\end{subequations}
where 
\begin{align}\label{eq:eta_def}
	\eta_{\rm d} \equiv \hidden{c^2 \times} \frac{1}{4 \pi \sigma} \approx \frac{\alpha \hidden{\hbar c^2}}{4\pi T}
\end{align}
is the magnetic diffusivity, assumed to be homogenous.  
\erefs{axmhdeq}{mhdeq} together with the Navier-Stokes equation for the plasma velocity ${\bf v}$ are the final equations to be solved. We first solve \eref{mhdeq} to determine the effect of the axion on the magnetic field, and afterward we will consider the evolution of the axion according to \eref{axmhdeq}.

\subsection{Effect of Axion on Magnetic Field}

In order to solve \eref{mhdeq} for the B-field we must know the fluid velocity, which appears in the advection term, ${\bm \nabla} \times ({\bf v} \times {\bf B})$.  
Since our primary interest is in the coevolution of the magnetic field and the axion, not in the magnetohydrodynamics, we will neglect this term \footnote{Strictly speaking the advective term is negligible compared to the diffusive term, $\eta_{\rm d} \nabla^2 {\bf B}$, only when $v < \eta_{\rm d} / \lambda_B \sim \hidden{\hbar c^2} (10^3 T \lambda_B)^{-1}$ where $\lambda_{B}$ is the typical length scale of the magnetic field.  Since $\lambda_{B} \gg T^{-1}$ is a macroscopic scale, this imposes a very strong upper bound on $v$, which is not easily satisfied.  We expect that inclusion of the advection term, perhaps in the context of a turbulent medium, will lead to a more rich and interesting solution, but that analysis is beyond the scope of this work.  }.  
Since \eref{mhdeq} is linear in ${\bf B}$, and we assume that $\ax$ is homogenous, we can solve the equation by first performing a Fourier transform.  
For a given mode ${\bf k}$ let $({\bf e}_1({\bf k}),{\bf e}_2({\bf k}),{\bf e}_3({\bf k}))$ form a right-handed, orthonormal triad of unit vectors with ${\bf e}_3({\bf k}) = {\bf k}/|{\bf k}|$. 
It is convenient to introduce the right- and left-circular polarization vectors by 
\begin{equation}
	{\bf e}^{\pm} ({\bf k}) = \frac{{\bf e}_1({\bf k}) \pm i {\bf e}_2({\bf k})}{\sqrt{2}} \per
\end{equation}
Note that $i {\bf k} \times {\bf e}^{\pm}({\bf k}) = \pm |{\bf k}| {\bf e}^{\pm}({\bf k})$.  
The mode decomposition is given by 
\begin{align}
	{\bf B}\tx = \int \! \! \frac{d^3 k}{(2 \pi)^3} e^{i{\bf k} \cdot {\bf x}} \sum_{{\rm s}=\pm} b_{\rm s}(t,|{\bf k}|) {\bf e}^{\rm s}({\bf k}) \per
\end{align}
With this replacement, \eref{mhdeq} becomes
\begin{align}
	\dot{b}_{\pm}(t,\kup) = - \eta_{\rm d} \kup^2 {b}_{\pm}(t,\kup) \pm \hidden{\frac{1}{c}} \gag \eta_{\rm d} \, \kup \, \dot{\ax} \, b_{\pm}(t,\kup)
\end{align}
where we have written $\kup = |{\bf k}|$.  
The last term of this equation, essentially the chiral-magnetic effect \cite{Vilenkin:1980fu}, has
been studied previously in the context of axions \cite{Field:1998hi} and cosmology 
\cite{Joyce:1997uy,Tashiro:2012mf}.
The solution is 
\begin{align}\label{bpmsoln}
	& b_{\pm} (t, \kup) = b_{\pm}(t_i, \kup) \ e^{-\kup^2 (t-t_i) \eta_{\rm d}} \ e^{\pm \kup / k_{\rm ax}(t)}
\end{align}
where we have defined the wavenumber
\be\label{eq:kax_def}
	k_{\rm ax}(t) \equiv \frac{2 \hidden{c}}{\gag \Delta \varphi(t) \eta_{\rm d}} 
\ee
and $\Delta \ax(t) = \ax(t) - \ax(t_i)$ is the change in the axion field.  
The prefactor in \eref{bpmsoln} is the initial spectrum of the magnetic field which will depend on the PMF generation mechanism.  
The first exponential is the usual diffusive decay term, which exponentially suppresses modes on a length scale shorter than $k_{\rm diff}^{-1} = \sqrt{(t-t_i) \eta_{\rm d}} \simeq 10^{-1} \sqrt{\hidden{\hbar c^2} t/T}$.  
The second exponential only kicks in at small length scales where $k > k_{\rm ax}$.  
Then it leads to a suppression of one polarization mode and enhancement of the other, depending on the sign of $\Delta \ax(t)$. 

The value of $\Delta \ax(t)$ depends on the solution for the axion field as well as the initial time $t_i$.  
We expect that the misalignment mechanism sets the initial condition $\ax(t_i) \sim \pm f_a$.  
The subsequent evolution is determined by solving \eref{axmhdeq}, which we will turn to in the next section.  
For the moment we will assume that the axion evolution is not significantly affected by the presence of the magnetic field, and the solution is the standard one:  the axion remains ``frozen'' at $\ax(t_i)$ until the time of the QCD phase transition when it begins to oscillate around $\ax = 0$ with angular frequency $\omega = m_a \hidden{c^2/\hbar}$ \cite{Kolb:1990}.  
Then we can approximate 
\be\label{eq:Dphiapprox}
	\Delta \varphi(t) \approx \begin{cases} 
	0 & t < t_{\QCD} \\
	{\rm s} f_{a} & t > t_{\QCD} 
	\end{cases}
\ee
where ${\rm s} = {\rm sign}[\varphi(t_i)]$.  
Using this approximation we can estimate $k_{\rm ax}$.  
Prior to the QCD phase transition, $\Delta \varphi$ is small and $k_{\rm ax}$ is large, meaning that none of the modes receive the enhancement or suppression from the axion coupling.  
This is reasonable since the axion is derivatively coupled, and as long as it is stationary there will be no effect on the magnetic field.  
After the QCD transition, we can estimate 
\begin{align}\label{eq:kax_estimate}
	k_{\rm ax} \approx \frac{4 \pi \sigma}{\gag f_{a} \hidden{c}} \approx \frac{4 \pi T}{\alpha \gag f_{a} \hidden{\hbar c}}
\end{align}
using \erefs{eq:sigma_def}{eq:eta_def}.  
Note that this result is insensitive to the Peccei-Quinn scale, and as long as $\gag = \alpha / (2\pi f_a)$ we have $k_{\rm ax}^{-1} \approx \alpha^2 / (8 \pi^2 T) \simeq 10^{-6}T^{-1} \hidden{\hbar c}$.

The solution in \eref{bpmsoln} can also be written as 
\be
	b_{\pm} (t, \kup) = b_{\pm}(t_i, \kup) \ e^{K^2 (t-t_i) \eta_{\rm d}} \ e^{- (\kup \mp K)^2 (t-t_i) \eta_{\rm d}}
\ee
where
\be\label{Kformula}
	K(t) 
	\equiv \frac{1}{k_{\rm ax} (t-t_{i}) \eta_{\rm d}}
	= \frac{\gag \Delta \ax(t)}{2 \hidden{c} (t-t_i)} \per
\ee
This representation of the solution is convenient, because all the spectral information is contained in the second factor.  
One of the helicity modes has a Gaussian spectrum peaked at $|K(t)|>0$ with width $\sqrt{1/(t-t_i) \eta_{\rm d}}$, and the other helicity mode peaks at $\kup = 0$.  
Estimating $k_{\rm ax}$ as above, we find that the associated length scale of the spectral peak corresponds to $K^{-1} \approx k_{\rm ax} t \eta_{\rm d} \simeq 600 \hidden{c} t$, which is larger than the scale of the cosmological horizon $d_{H} \sim \hidden{c} t$.  

It appears that the presence of an axion condensate coupled to electromagnetism has a negligible impact on the evolution of a primordial magnetic field, unless there are situations in which $\Delta \ax$ can be much larger than $f_{a}$.  

We note that our analysis ignores the possibility of turbulence in the primordial plasma. It
would be of interest to include both turbulence and the axion coupling in future studies.

\subsection{Effect of Magnetic Field on Axion}

Next we will investigate the effect of a background magnetic field on the axion condensate.  
We have seen that the magnetic field is approximately unmodified on length scales larger than the diffusion length, $k_{\rm diff}^{-1} \sim \sqrt{\eta_{\rm d} t}$ ({\it cf.}, \eref{bpmsoln}).  
In this regime \eref{axmhdeq} can be rewritten as 
\begin{align}\label{eq:a_reduced}
	& \ddot{\ax} + 2 \frac{\dot{\ax}}{\tau_{\rm decay}} + \frac{\ax}{\tau_a^2}  = \Hcal
\end{align}
where
\begin{align}
&  \tau_{\rm decay} \equiv \frac{8 \pi}{\gag^2 \eta_{\rm d} \langle |{\bf B}|^2\rangle } \com \\
& \tau_{a} \equiv \hidden{\frac{\hbar}{c^2}} \frac{1}{m_a} \com \\
& \Hcal \equiv \hidden{c} \frac{\gag \eta_{\rm d}}{4 \pi} \langle {\bf B} \cdot {\bm \nabla} \times {\bf B} \rangle ,
\end{align}
and the angled brackets $\langle \cdot \rangle$ denote spatial averaging.  
The axion condensate evolves like a damped and driven harmonic oscillator, where the damping and driving forces are induced by the magnetic field background.  
As we discuss below, it is interesting that the driving force is associated with the {\it helicity} of the magnetic field.

The magnetic-induced damping of axion oscillations is parametrized by the time scale $\tau_{\rm decay}$.  
To determine when this damping will be relevant for the evolution of the axion, we compare it with the cosmological time scale, given by \eref{eq:tautH}.  
To express $\langle |{\bf B}|^2\rangle = B^2$ in terms of the magnetic field strength today, $B_0$, we use $B = B_0 (a_0 / a)^2 \simeq 10 B_0 (T / T_0)^2$ 
where the factor of $10$ is related to the number of relativistic degrees of freedom in the early universe and today.   
Then the ratio is found to be 
\begin{align}
	\frac{\tau_{\rm decay}}{t_H} 
	& \approx \frac{16 \pi^2 g_{\ast}}{75} \frac{T_0^4}{\alpha \gag^2 M_P B_0^2 T} \hidden{ \frac{1}{\hbar^2 c^4} } \\
	& \simeq 10^{16} \frac{(10^{-13} \GeV^{-1} \hidden{\sqrt{\hbar c}})^2}{\gag^2} \frac{(10^{10} \GeV)}{T} \frac{(10^{-14} \, {\rm G})^2}{B_{0}^2} \per \nonumber
\end{align}
This estimate suggests that the magnetic-induced decay of the axion field is negligible for a typical Peccei-Quinn scale and B-field strength.  
If the B-field strength today were as large as $B_{0} \sim 10^{-9} \Gauss$ and the Peccei-Quinn scale was as low as $f_{a} \sim \TeV \hidden{\sqrt{\hbar c}}$, then $\tau_{\rm decay}$ would be comparable to the Hubble time at $T \approx f_{a} \hidden{\sqrt{\hbar c}}$.  
As the temperature decreases, the magnetic-induced decay becomes less relevant.  

It is interesting that the magnetic field also induces a driving force, parametrized by $\Hcal$.  
The pseudoscalar $\Hcal$ is related to the helicity of the magnetic field.  
This is perhaps more evident from the initial form of the axion field equation, \eref{axeq}, where ${\bf E} \cdot {\bf B}$ is equal to the rate of change of the helicity density $-(1/2\hidden{c}) d( {\bf A} \cdot {\bf B})/dt$ plus a divergence, which vanishes upon spatial averaging.  
If the power in the magnetic field is localized on a particular length scale $\lambda_B$ we can estimate 
$\langle {\bf B} \cdot {\bm \nabla} \times {\bf B} \rangle \sim B^2 / \lambda_B \sim 300 (B_0^2 / \lambda_{B_0})(T / T_0)^5$ where we used $\lambda_{B} \sim 3 \lambda_{B,0} (T_0/T)$. 

Prior to the QCD phase transition we can neglect the mass and drag terms in \eref{eq:a_reduced}, and the solution is simply $\ax = \Hcal t^2 / 2$.  
Since the axion is massless, there is no restorative potential, and the helical magnetic field leads to an unbounded growth of the axion condensate.  
Although this analysis neglects the Hubble drag, we can estimate the maximum field excursion in one one Hubble time to be $\Hcal t_H^2 / 2$.  
Comparing with the Peccei-Quinn breaking scale, the corresponding angular excursion is 
\begin{align}
	\Delta \theta 
	& \approx \frac{\Hcal t_H^2}{f_a}
	\approx \frac{75 \hidden{\hbar^3 c^7} }{16 \pi^2} \frac{\alpha \gag M_P^2 B_0^2}{f_a g_{\ast}^2 T_0^5 \lambda_{B,0}} \nn
	& \simeq 10^{-35} 
	\left( \frac{B_0}{10^{-14} \, {\rm G}} \right)^{2} 
	\left( \frac{\lambda_{B,0}}{10 \Mpc} \right)^{-1} 
	\nn & \qquad \times 
	\left( \frac{\gag}{10^{-13} \GeV^{-1} \hidden{\sqrt{\hbar c}}} \right) 
	\left( \frac{f_{a}}{10^{10} \GeV \hidden{/\sqrt{\hbar c}}} \right)^{-1} 
	\per 
\end{align}
We are led to conclude that for realistic parameters, the helical PMF does not significantly impact the evolution of the axion condensate prior to the QCD phase transition.  

After the QCD phase transition, the axion mass reaches its asymptotic value, and the source term displaces the minimum of the axion potential from $\ax = 0$ to $\ax_{\rm min} = \Hcal \tau_{a}^2 = \hidden{\hbar^{2}} \Hcal / m_{a}^{2} \hidden{c^4}$.  
In terms of the angular coordinate:
\begin{align}\label{eq:theta_min}
	\theta_{\rm min} 
	& \approx \frac{\Hcal \tau_{a}^2}{f_a}
	\approx \frac{25 \hidden{\hbar^3}}{12\pi^2 \hidden{c}} \frac{\alpha \gag B_0^2 T^4}{f_a m_a^2 \lambda_{B,0} T_{0}^5} \\
	& \simeq 10^{-47} 
	\left( \frac{B_0}{10^{-14} \Gauss} \right)^2 
	\left( \frac{\lambda_{B,0}}{10 \Mpc} \right)^{-1} 
	\left( \frac{T}{200 \MeV} \right)^4 \nn
	& \quad \times
	\left( \frac{f_{a}}{10^{10} \GeV \hidden{/\sqrt{\hbar c}}} \right)^{-1} 
	\left( \frac{\gag}{10^{-13} \GeV^{-1} \hidden{\sqrt{\hbar c}}} \right) 
	\left( \frac{m_{a}}{1 \meV} \right)^{-2} \nonumber \per
\end{align}
The temperature dependence enters through $B \sim T^2$, $\lambda_{B} \sim 1 / T$, and $\eta_{\rm d} \sim 1 / T$, and the fractional shift is largest at high temperature where $B$ is large and $\lambda_{B}$ is small.  
Immediately after the QCD phase transition, $T_{\QCD} \sim 200 \MeV$, the fractional shift is already extremely small.  
Moreover if the PMF is not helical then $\Hcal = 0$ and there is no shift in the axion potential.  

It is interesting that the estimate of \eref{eq:theta_min} is insensitive to the Peccei-Quinn scale; as long as $\gag = \alpha / (2\pi f_a)$ and $m_a = \Lambda_{\QCD}^2 / f_a$ we have $\gag / f_a m_a^2 = \alpha / (2 \pi \Lambda_{\QCD}^4)$.  
Then the primarily challenge toward obtaining a large effect is the smallness of the magnetic field strength.  
Although unrelated to primordial magnetic fields, which is the motivation for this work, it would be interesting to study the axion condensate in an astrophysical system where the magnetic field is both helical and strong.
For instance, the field strength in a magnetar can grow as large as $B \sim 10^{15} \Gauss$ and
the magnetic field in some astrophysical jets is known to be helical \cite{2012SSRv..169...27P}.

\subsection{Axion-Photon Interconversion}

Until this point we have focused our attention on the interplay between the axion condensate and the primordial magnetic field, and we now turn our attention to the quanta of these fields. 
In the presence of a background magnetic field, the interaction in \eref{axLag} yields a mixing between axion particles and photons.  
Typically the conversion is inefficient, but in the presence of a plasma the photon acquires an effective mass, and the conversion probability experiences a resonance when $m_{\gamma} = m_{a}$ \cite{Yanagida:1987nf}.  
In the cosmological context, the conversion of photons into axions may lead to a dimming of the cosmic microwave background across frequencies.  
Then measurements of the spectrum of the CMB can be used to place constraints on the axion-photon coupling and the magnetic field strength.  

Bounds were obtained from the COBE / FIRAS measurement of the CMB spectrum in \rref{Mirizzi:2005ng}, and recently a second group \cite{Tashiro:2013yea} has extended the calculation to include forecasts for next-generation CMB telescopes, namely PIXIE and PRISM.  
The latter references finds an upper bound on the product of the axion-photon coupling and the {\rm r.m.s.} magnetic field strength today:  
\begin{align}\label{eq:gB_constraint}
	\gag B_0 < 10^{-14} \GeV^{-1} \, {\rm nG} \quad & \text{(COBE-FIRAS data)} \nn
	\gag B_0 < 10^{-16} \GeV^{-1} \, {\rm nG} \quad & \text{(PIXIE/PRISM forecast)} 
\end{align}
for a light axion $m_{a} < 10^{-14} \eV$.  
For larger axions masses the bound weakens.  
Using our fiducial value for the magnetic field $B_0 \simeq 10^{-14} \Gauss$, we can write 
\begin{align}\label{eq:gB_constraint_2}
	\gag < 10^{-9} \GeV^{-1} \, \left( \frac{B_0}{10^{-14} \Gauss} \right)^{-1} \qquad & \text{(COBE-FIRAS)} \nn
	\gag < 10^{-11} \GeV^{-1} \, \left( \frac{B_0}{10^{-14} \Gauss} \right)^{-1} \qquad & \text{(PIXIE/PRISM)} \per
\end{align}
These bounds are comparable to the direct search limits from the CAST helioscope \cite{Andriamonje:2007ew}, 
\begin{align}\label{eq:CAST}
	\gag < 8.8 \times 10^{-11} \GeV^{-1} 
\end{align}
for $m_{a} \lesssim 0.02 \eV$.  

\section{Dirac Neutrinos}
\label{sec:neutrinos}

While the neutrinos are known to be massive particles, the nature of their mass remains a mystery.  
If neutrinos are Dirac particles then the theory contains four light states per generation: an active neutrino $\nu_{L}$, an active antineutrino, $\bar{\nu}_{R}$, a sterile neutrino $\nu_{R}$, and a sterile antineutrino $\bar{\nu}_{L}$.  
The active states interact through the weak force, and this allows them to come into thermal equilibrium in the early universe.  
The sterile states, on the other hand, interact only via the Yukawa interaction with the Higgs boson, and because of the smallness of the Yukawa coupling $y_{\nu} \sim m_{\nu} / v \sim 10^{-12}$, these states are not expected to be populated.  

This story is modified if a strong magnetic field permeated the early universe.  
The nonzero neutrino mass implies that the neutrino will also have a nonzero magnetic moment $\mu_{\nu}$.  
From Standard Model physics alone one expects \cite{Lee:1977tib, Fujikawa:1980yx}
\begin{align}\label{eq:mnu_SM}
	\mu_{\nu}^{\text{\sc sm}} 
	\simeq (3\times 10^{-20} \mu_{B}) \frac{m_{\nu}}{0.1 \eV \hidden{/c^2}} \com
\end{align}
where $\mu_B \equiv e \hidden{\hbar}/ 2m_e \hidden{c} \simeq 83.6 \GeV^{-1} \hidden{(\hbar c)^{3/2}} $ is the Bohr magneton, but new physics can increase this value appreciably.  
The magnetic field couples to $\mu_{\nu}$ and induces the spin-flip transitions $\nu_{L} \to \nu_{R}$ and $\bar{\nu}_{R} \to \bar{\nu}_{L}$, which can be viewed as the absorption or emission of a photon.
If the spin-flip occurs rapidly in the early universe, the sterile states would be populated, and the effective number of relativistic neutrino species would double from $N_{\nu} = 3$ to $6$.  
However, this is not consistent with measured abundances of the light elements, which imply $N_{\nu} \approx 3$ at the time of nucleosynthesis \cite{Izotov:2010ca}.  
We must therefore require that the spin-flip transition goes out of equilibrium prior to the QCD epoch, $T_{\text{\sc qcd}} \simeq 200 \MeV$, so that the subsequent entropy injection at the QCD phase transition can suppress the relative abundance of sterile states to acceptable levels \cite{Kolb:1990}.  
This translates into an upper bound on the neutrino magnetic moment and magnetic field strength, which was originally discussed by Enqvist et al. \cite{Enqvist:1992di,Enqvist:1994mb}.  

In the rest of this section we apply the results of \rref{Enqvist:1994mb}.  
The spin-flip transition occurs with a rate
\begin{align}
	\Gamma_{L \to R} = \langle P_{\nu_{L} \to \nu_{R}} \rangle \Gamma_{W}^{\rm tot}
\end{align}
where $\langle P_{\nu_{L} \to \nu_{R}} \rangle$ is the average conversion probability and $\Gamma_{W}^{\rm tot}$ is the total weak scattering rate.  
The active neutrinos scatter via the weak interaction which leads to $\Gamma_{W}^{\rm tot} \simeq 30 G_{F}^2 T_{\text{\sc qcd}}^5 \hidden{\hbar^{-7} c^{-6}}$ at the QCD epoch.  
The conversion probability depends on the magnetic moment and field strength as $\langle P_{\nu_{L} \to \nu_{R}} \rangle \propto \mu_{\nu}^2 B^2$, since the interaction Hamiltonian is $H_{\rm int} = - {\bm \mu}_{\nu} \cdot {\bf B}$.  
The coefficient takes different values depending on the relative scale of the magnetic field domains $\lambda_{B}$ and the weak collision length $L_{W} \approx \hidden{c} (\Gamma_{W}^{\rm tot})^{-1}$.  
At the QCD epoch $L_{W} \simeq 1.6 \times 10^{-2} \cm$, which corresponds to a length scale of $L_{W,0} \simeq 3 \times 10^{10} \cm$ today.  
It is safe to assume that the magnetic field of interest is much larger than this length scale, and therefore $\lambda_{B} \gg L_{W}$.  
To ensure that the spin-flip transition is out of equilibrium one must impose $\Gamma_{L\to R} < H$ with $H$ the Hubble parameter.  
This inequality resolves to the bound (see Eq.~(37) of \rref{Enqvist:1994mb})
\begin{align}
	\mu_{\nu} B(t_{\QCD}) < (3.5 \times 10^{2} \mu_{B} \Gauss) \sqrt{\frac{L_{W}}{\lambda_{B}}} \per
\end{align}
To express this inequality in terms of the B-field strength and correlation length today, we use $B \simeq 6 B_{0} ( T_{\QCD} / T_{0} )^2$ and $L_{W} / \lambda_{B} = L_{W,0} / \lambda_{B,0}$.  
This leads to an upper bound on the neutrino magnetic moment: 
\begin{align}\label{eq:mnu_Bbound}
	\mu_{\nu}  < (3 \times 10^{-16} \mu_{B}) \left( \frac{B_{0}}{10^{-14} \Gauss} \right)^{-1} \left( \frac{\lambda_{B,0}}{10 \Mpc} \right)^{-1/2} \per
\end{align}
If this bound is not satisfied, the sterile neutrino states will still be thermalized with the active neutrino states at the time of BBN leading to $N_{\nu} \approx 6$, which is inconsistent with the data.  
If the neutrinos are Majorana particles, then the sterile states are much heavier, and this bound does not apply.  

 \begin{figure}[t]
\hspace{0pt}
\vspace{-0in}
\begin{center}
\includegraphics[width=0.47\textwidth]{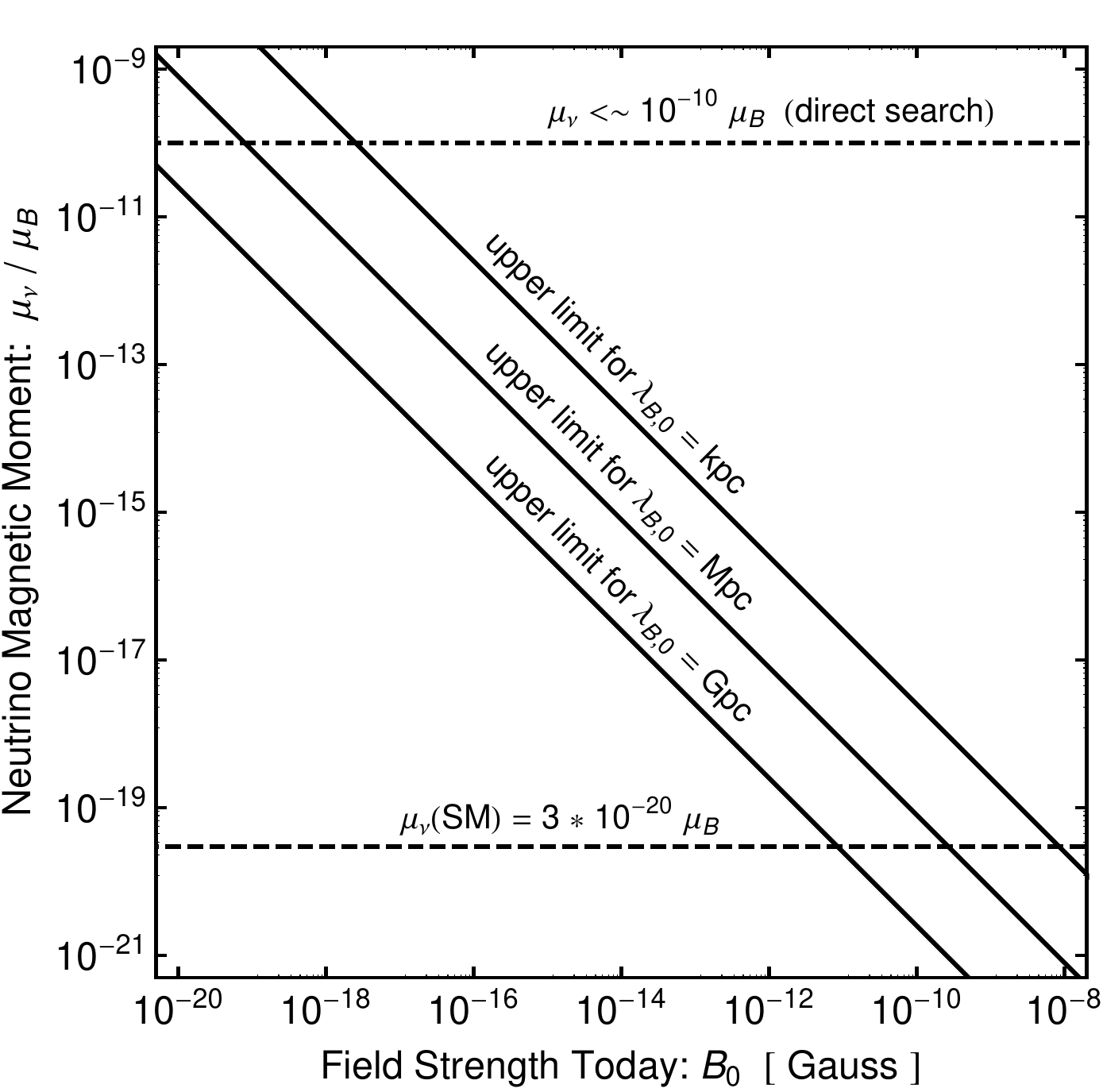} 
\caption{
\label{fig:mag_mom}
The requirement that spin-flip transitions are out of equilibrium at the QCD epoch leads to an upper bound the neutrino magnetic moment given by \eref{eq:mnu_Bbound}.  
}
\end{center}
\end{figure}

The bound in \eref{eq:mnu_Bbound} is represented graphically in \fref{fig:mag_mom}.  
For comparison we show the SM prediction from \eref{eq:mnu_SM} and the direct search limits.  
The strongest laboratory constraints arise from elastic $\nu-e$ scattering.  
The limits are flavor-dependent, but they are typically at the level of \cite{Agashe:2014kda}
\begin{align}\label{eq:mnu_direct}
	\mu_{\nu} \lesssim 10^{-10} \mu_{B}
	\qquad \text{(direct)} \per
\end{align}
From the figure we see that the indirect early universe constraint is significantly stronger than the direct constraint for $B \gtrsim 10^{-18} \Gauss$.  
This provides the exciting opportunity to constrain extensions of the SM that predict an enhancement to the magnetic moment of Dirac neutrinos.

\section{Summary}
\label{conclusion}

Growing evidence for the existence of an intergalactic magnetic field has motivated us to consider the effects of a primordial magnetic field on models of exotic particle physics in the early universe.  
We have focused our study on magnetic monopoles, axions, and Dirac neutrinos with a magnetic moment.  
We summarize our results here.  

In the context of a universe containing relic magnetic monopoles, we have derived a ``primordial Parker bound'' by requiring the survival of a primordial magnetic field until the time of electron-positron annihilation.  
The bound, which appears in \eref{eq:n0_bound}, gives an upper limit on the cosmological monopole number density today: $n_{0} < 1 \times 10^{-20} \cm^{-3}$.  
This translates into an upper bound on the monopole flux in the Milky Way; if the monopoles are unclustered then $\Fcal < 3 \times 10^{-14} \cm^{-2} \sec^{-1} \sr^{-1} (v / 10^{-3} \hidden{c})$, and if they are clustered the bound weakens by a factor of $\sim 10^{5}$.  
In \fref{fig:constraints} we compare the primordial Parker bound with other constraints on relic monopoles.  
If the primordial magnetic field is not generated prior to $T \simeq \MeV$, then this bound does not apply.  

After $e^+ e^-$ annihilation the monopoles are able to free stream, and they evolve along with the magnetic field as described by the system of equations in \eref{eq:coevolution}.  
The solution is an analog of the familiar plasma oscillations (``Langmuir oscillations'') seen in an electron-ion plasma.  
In the regime where the plasma oscillations are fast compared to the cosmological expansion, the coupling of the monopoles to the magnetic field leads to an anomalous scaling with redshift such that $B \sim a^{-9/4}$, $v_{m} \sim a^{-3/4}$, and $\rho_{B} \sim \rho_{\rm kin.} \sim a^{-9/2}$.  
The behavior of the coupled system is effectively the average of the usual scalings for radiation $\rho_{B} \sim a^{-4}$ and the kinetic energy of a non-relativistic gas $\rho_{\rm kin.} \sim a^{-5}$.  
If the strength of the intergalactic magnetic field could be measured over a range of redshifts, this would allow for a direct test of the anomalous scaling, and thereby probe relic magnetic monopoles.  

We have also studied the effect of a primordial magnetic field on the evolution of an axion condensate in the early universe.  
We obtain an exact solution to the MHD equations for the magnetic field in the limit where the advection term is negligible and the axion is homogenous.  
After Peccei-Quinn breaking but prior to the QCD phase transition, the axion field is frozen, because its mass is smaller than the Hubble scale, and since the axion is derivatively coupled, this leads to no effect on the magnetic field.  
Below the QCD scale the axion field begins to oscillate, and the spectrum of the magnetic field is distorted as in \eref{bpmsoln}.  
One helicity mode of the magnetic field is enhanced while the other is suppressed; this CP-violation is a consequence of the axion's pseudoscalar nature.  
However, the spectral shape of the magnetic field is only affected on extremely large
length scales, as given by \eref{Kformula}, except in situations where there can be significant axion evolution prior 
to the QCD epoch.

We next study the evolution of the homogenous axion condensate in the presence of a background magnetic field.  
The axion behaves as a damped and driven harmonic oscillator, as seen from its equation of motion \eref{eq:a_reduced}.  
The damping time scale depends on the strength of the magnetic field and the photon-axion coupling.  
For typical parameters it is generally larger than the cosmological time scale, and therefore irrelevant for the evolution of the axion.  
It is interesting that the driving force (source term $\Hcal$) is only operative when the magnetic field has a helicity.  
This can be seen directly from the interaction $\Lcal \ni \ax F \widetilde{F}$ where $F \widetilde{F} \sim {\bf E} \cdot {\bf B} \sim \dot{h}$ is related to the rate of change of magnetic helicity $h = {\bf A} \cdot {\bf B}$. 
Prior to the QCD phase transition when the axion was effectively massless, the axion field equation reduces to $\ddot{\ax} = \Hcal$.  
In principle a very strong magnetic field could cause the axion to grow without bound as $\ax(t) = \Hcal t^2/2$, by drawing energy from the magnetic field.  
For typical parameters, however, this growth occurs on a time scale that is much longer than the cosmological time.
It may still be the case that helical magnetic fields occurring in astrophysical environments are strong enough to lead to observable signatures.  

We have also considered the resonant conversion of CMB photons into axions, which leads to a distortion of the CMB blackbody spectrum \cite{Yanagida:1987nf, Mirizzi:2005ng, Tashiro:2013yea}.  
Using constraints on spectral distortions from current and anticipated future CMB telescopes, \rref{Tashiro:2013yea} obtained an upper bound on the axion-photon coupling. For our fiducial magnetic field strength this translates
into $\gag \lesssim 10^{-9} \GeV^{-1}$ with current data, and a forecast of $\gag \lesssim 10^{-11} \GeV^{-1}$
for experiments presently under discussion (see \eref{eq:gB_constraint_2}).

Finally we turn to the effect of the primordial magnetic field 
on Dirac neutrinos, which carry a magnetic moment.  
In the presence of a magnetic field, left-handed neutrinos can be converted into right-handed neutrinos.  
If this spin-flip process is in equilibrium in the early universe, the right-handed states would be populated, and the effective number of relativistic neutrino species would double from $3$ to $6$, which is inconsistent with observations.  
Requiring that this process is out of equilibrium at the time of the QCD phase transition leads to an upper bound on the neutrino magnetic moment and magnetic field strength.  
Drawing on the work of \rref{Enqvist:1994mb}, we find the limit in \eref{eq:mnu_Bbound}, which implies $\mu_{\nu} < 3 \times 10^{-16} \mu_{B}$ for our fiducial magnetic field parameters $B_{0} = 10^{-14} \Gauss$ and $\lambda_{B} = 10 \Mpc$. As seen in \fref{fig:mag_mom}, this bound is significantly stronger than the direct search limits over most of the parameter space.

\acknowledgments
We are grateful to David Marsh and Hiroyuki Tashiro for comments.
TV gratefully acknowledges the Clark Way Harrison Professorship at Washington University 
during the course of this work. This work was supported by the DOE at ASU. 
A.J.L. was also supported in part by the National Science Foundation under grant number PHY-1205745.

\bibstyle{aps}
\bibliography{parkerbound}

\end{document}